\documentclass[twocolumn,aps,pra,showpacs,floatfix,superscriptaddress]{revtex4-1}

\usepackage[english]{babel}
\usepackage{epsfig}
\usepackage{pst-pdf}
\usepackage{graphicx}
\usepackage{amsmath}
\usepackage{subfigure}
\usepackage{eso-pic}   
\usepackage{rotating}  
\usepackage{graphicx}
\usepackage{hyperref}
\usepackage{dcolumn}
\usepackage{amssymb}
\usepackage{color}
\usepackage{verbatim,moreverb,bm}
\def\sss{\scriptscriptstyle\rm}

\newcommand{\be}{\begin{equation}}
\newcommand{\ee}{\end{equation}}
\newcommand{\bea}{\begin{eqnarray}}
\newcommand{\eea}{\end{eqnarray}}

\def\UP{\uparrow}
\def\DN{\downarrow}
\def\nn{\nonumber}

\def \E {\mathcal{E}}
\def \hxc {\text{Hxc~}}
\def \SE {Schr\"odinger equation~}

\def \fig{Fig.~}

\begin{document}
\title{The time-dependent exchange-correlation functional for a Hubbard dimer: quantifying non-adiabatic effects}

\author{Johanna I. Fuks$^*$}
\affiliation{Nano-Bio Spectroscopy group and ETSF, Dpto.~F\'isica de Materiales, Universidad del Pa\'is Vasco, Centro de
F\'isica de Materiales CSIC-UPV/EHU-MPC and DIPC, Av.~Tolosa 72, E-20018 San 
Sebasti\'an, Spain}
\affiliation{Department of Physics and Astronomy, 
Hunter College and the Graduate Center of the City University of New York, New York City, United States}

\author{Mehdi Farzanehpour$^*$}
\affiliation{Nano-Bio Spectroscopy group and ETSF, Dpto.~F\'isica de Materiales, Universidad del Pa\'is Vasco, Centro de
F\'isica de Materiales CSIC-UPV/EHU-MPC and DIPC, Av.~Tolosa 72, E-20018 San 
Sebasti\'an, Spain}
\thanks{These two authors contributed equally}

\author{Ilya V. Tokatly}
\affiliation{Nano-Bio Spectroscopy group and ETSF, 
Dpto.~F\'isica de Materiales, Universidad del Pa\'is Vasco, Centro de
F\'isica de Materiales CSIC-UPV/EHU-MPC and DIPC, Av.~Tolosa 72, E-20018 San 
Sebasti\'an, Spain}
\affiliation{IKERBASQUE, Basque Foundation for Science, E-48011 Bilbao, Spain}

\author{Heiko Appel}
\affiliation{Fritz-Haber-Institut der Max-Planck-Gesellschaft, Faradayweg 4-6,
D-14195 Berlin, Germany}

\author{Stefan Kurth}
\affiliation{Nano-Bio Spectroscopy group and ETSF, 
Dpto.~F\'isica de Materiales, Universidad del Pa\'is Vasco, Centro de
F\'isica de Materiales CSIC-UPV/EHU-MPC and DIPC, Av.~Tolosa 72, E-20018 San 
Sebasti\'an, Spain}
\affiliation{IKERBASQUE, Basque Foundation for Science, E-48011 Bilbao, Spain}

\author{Angel Rubio}
\affiliation{Nano-Bio Spectroscopy group and ETSF, 
Dpto.~F\'isica de Materiales, Universidad del Pa\'is Vasco, Centro de
F\'isica de Materiales CSIC-UPV/EHU-MPC and DIPC, Av.~Tolosa 72, E-20018 San 
Sebasti\'an, Spain}
\affiliation{Fritz-Haber-Institut der Max-Planck-Gesellschaft, Faradayweg 4-6,
D-14195 Berlin, Germany}

\date{\today}

\begin{abstract}
We address and quantify the role of non-adiabaticity ("memory 
effects") in the exchange-correlation (xc) functional of time-dependent 
density functional theory (TDDFT) for describing non-linear dynamics of 
many-body systems. Time-dependent resonant processes are particularly 
challenging for available TDDFT approximations, due to their strong non-linear and non-adiabatic 
character. None of the known approximate density functionals are able to cope 
with this class of problems in a satisfactory manner.
In this work we look at the prototypical example of the resonant processes by 
considering Rabi oscillations within the exactly soluble 2-site Hubbard model. 
We construct the exact adiabatic xc functional  and show that (i) it 
does not reproduce correctly resonant Rabi dynamics, (ii) there is a sizable 
non-adiabatic contribution to the exact xc potential, which turns out to be 
small only at the beginning and at the end of the Rabi cycle when the 
ground state population is dominant. We then propose a "two-level" 
approximation for the time-dependent xc potential which can capture Rabi 
dynamics in the 2-site problem.
It works well both for 
resonant and for detuned Rabi oscillations and becomes essentially exact in 
the linear response regime.  This new, fully non-adiabatic and explicit 
density functional constitutes one of the main results of the present work.

\end{abstract} 

\maketitle


\section{Introduction}
Due to the favorable balance between efficiency and accuracy,  
time-dependent density functional theory (TDDFT) is becoming the theory of 
choice to describe the interaction of many-electron systems with external 
electromagnetic fields of arbitrary intensity, shape and time dependence. 
Within this theory the observables are expressible as functionals of the 
time-dependent density. Similarly to static DFT, in TDDFT one can define an 
auxiliary non-interactiong Kohn-Sham (KS) system which reproduces the exact 
time-dependent dynamics of the density. It is the propagation of this 
auxiliary system in an (unknown) effective local potential which makes TDDFT 
computationally powerful. However, despite of the great success of the theory 
in describing optical properties of a large variety of molecules and 
nanostructures \cite{TDDFTBook2012,Casida09, LB07}, the available 
approximations for the exchange-correlation (xc) potential 
exhibit serious deficiencies in the description of
non-linear processes,  long range charge transfer \cite{M05,GB04, T03} and 
double excitations \cite{MZCB2004,HFTAGR11, EGCM11}, to mention a few. 

The theoretical challenge is to improve the available functionals in order to 
capture the nonlocality both in space and time of the exact xc functional 
which depends on the entire history of the density, the 
initial (interacting) many-body state and the initial KS state 
\cite{FarTok2012,EFRM12,MBW02,EM12,RNvL13}.

We note that almost all TDDFT calculations today use an adiabatic 
approximation for the xc potential, which uses the instantaneous density 
as input for an approximate ground-state functional. Thus, this approximation 
completely neglects both the history and the initial-state dependence of 
the exact functional. 

The successes and failures of the adiabatic approximation to describe linear 
response phenomena have been addressed in many works 
\cite{TDDFTBook2012,Ullrich:12, Casida09, KK09, BKK11}.
However, much less is 
known about the performance of adiabatic TDDFT for general dynamics beyond 
linear response. In some of our past studies 
\cite{FHTR11, HFTAGR11,EFRM12,FERM13} on one-dimensional model systems we 
have shown that adiabatic xc functionals fail to describe dynamical processes 
where the density changes significantly in time (e.~g. in photo-physical and 
chemical processes where valence electrons are promoted to empty states).
There are few cases where the exact time-dependent xc potential is known and 
can thus be used to test approximations  \cite{TGK08, RG12, EFRM12}.
In these works it has been shown numerically that novel dynamical 
steps appear in the xc potential which are fundamental to capture the proper 
resonant versus non-resonant dynamics and charge localization.  
While the construction of accurate approximations to the 
exact universal xc functional of TDDFT for Coulomb systems remains a challenge, 
simple model Hamiltonians constitute a convenient framework to gain insights 
into the properties of the exact TDDFT functional. 

In the present work we exploit the possibilities of a solvable  lattice 
model  -- the 2-site Hubbard model ~\cite{CF12, U06, V08}  
-- to address the impact 
of non-locality in time in the exchange correlation functional of TDDFT.  
Specifically, we study resonant Rabi oscillations, a prototypical example of 
non-linear external field driven dynamics where the population of states 
changes dramatically in time. We first derive here the exact ground-state 
Hartree-exchange-correlation (Hxc) functional for the 2-site model
using the Levy-Lieb constrained search\cite{Levy:79,Levy82,Lieb:83}.  
This functional, when used in a TDDFT context with the 
instantaneous time-dependent density as input, constitutes the exact 
adiabatic approximation which can be used as a reference to quantify the role 
of memory effects. By carefully studying and quantifying the dynamics produced 
by TDDFT with the adiabatic Hxc potential we demonstrate that it fails both 
quantitatively and qualitatively to describe Rabi oscillations. In the second 
part of this work we apply an analytic density-potential map for lattice 
systems \cite{LiUll2008,FarTok2012} to derive an explicit, fully non-adiabatic 
xc density functional which correctly captures all features of Rabi dynamics 
in the Hubbard dimer. This functional is one of the main results of this paper.

The paper is organized as follows: in Sec.~\ref{s:HM} we introduce the physics 
of the Rabi effect for the Hubbard dimer, showing how the dipole moment and 
state occupations evolve with time during the course of resonant Rabi 
oscillations. 
In Sec.~\ref{s:ksHM} we address the same problem from a TDDFT perspective. In 
particular we use the exact adiabatic xc functional as a reference to quantify 
memory effects. In the Sec.~\ref{s:exhxcF} we consider the exact interacting 
system in a two-level approximation which allows us to derive a new 
approximate Hxc potential as an explicit functional of the time-dependent 
density. The excellent performance of this approximation is demonstrated and 
explained. We end the paper with our conclusions in Sec.~\ref{s:conclus}. In 
the Appendix we derive the exact ground state xc potential for the Hubbard 
dimer using the Levy-Lieb constrained search.


\section{Rabi oscillations for two-site Hubbard model} 
\label{s:HM}

We consider the dynamics of two electrons on a Hubbard dimer, that is, a 
two-site interacting Hubbard model with on-site repulsion $U$ and hopping 
parameter $T$. The Hamiltonian of the system reads
\begin{align}
 \hat{H}= & -T \sum_\sigma \left( \hat{c}_{1\sigma}^\dag \hat{c}_{2\sigma} +\hat{c}_{2\sigma}^\dag\hat{c}_{1\sigma} \right)
+ U \left( \hat{n}_{1 \uparrow} \hat{n}_{1 \downarrow} + \hat{n}_{2 \uparrow} \hat{n}_{2 \downarrow}\right)  \nonumber \\ 
          & + \sum_{\sigma} \left( v_1(t) \hat{n}_{1 \sigma} + v_2 (t) \hat{n}_{2 \sigma} \right),
\label{eq:HubbardH}
\end{align}
where $\hat{c}_{i\sigma}^{\dag}$ and $\hat{c}_{i\sigma}$ are creation and 
annihilation operators for a spin-$\sigma$ electron on site $i$, respectively. 
The $\hat{n}_{i\sigma}=\hat{c}_{i\sigma}^{\dag} \hat{c}_{i\sigma}$ are the operators 
for the spin-$\sigma$ density at site $i$, and the $v_{1,2}(t)$ are 
time-dependent on-site potentials. We use $\hbar=e=1$ throughout this work.  Energies are given in units of the hopping parameter $T$.
As we will see, this simple model captures most qualitative features of Rabi 
oscillations in interacting systems as well as the main difficulties of 
describing Rabi dynamics within TDDFT.

The many-body time-dependent Schr\"odinger equation,
\begin{equation}\label{eq:SE}
  i \partial_t |\psi(t)\rangle = H(t)|\psi(t)\rangle ,
\end{equation}   
describes the evolution of the system from a given initial state 
$|\psi_0\rangle$. Since the Hamiltonian \eqref{eq:HubbardH} is independent of 
spin, the spin structure of the wave function $|\psi(t)\rangle$ is fixed by 
the initial state. In the following we study the evolution from the ground 
state of the Hubbard dimer and therefore it is sufficient to consider only the 
singlet sector of our model.

In the absence of an external potential, $v_{1,2}=0$, the stationary singlet 
eigenstates of the Hamiltonian \eqref{eq:HubbardH} take the form
\begin{subequations}\label{eq:states}
\begin{eqnarray}  
 && \begin{split} \label{eq:GS}
   |g\rangle =& \:\mathcal{N}_g \Big( \hat c_{1 \UP} ^\dag \hat c_{1 \DN} ^\dag + \hat c_{2 \UP} ^\dag \hat c_{2 \DN} ^\dag + \beta_+ \big(\hat c_{1 \UP} ^\dag \hat c_{2 \DN} ^\dag-\hat c_{1 \DN} ^\dag \hat c_{2 \UP} ^\dag \big) \Big) |0\rangle 
  \end{split},\\
  &&|e_1 \rangle = \: 1/ \sqrt{2} \big( \hat c_{1 \UP} ^\dag \hat c_{1 \DN} ^\dag + \hat c_{2 \UP} ^\dag \hat c_{2 \DN} ^\dag\big) |0\rangle,\\
  &&\begin{split}
   |e_2\rangle = \:&\mathcal{N}_{e_2} \Big( \hat c_{1 \UP} ^\dag \hat c_{1 \DN} ^\dag + \hat c_{2 \UP} ^\dag \hat c_{2 \DN} ^\dag  + \beta_- \big(\hat c_{1 \UP} ^\dag \hat c_{2 \DN} ^\dag-\hat c_{1 \DN} ^\dag \hat c_{2 \UP} ^\dag \big) \Big) |0\rangle, \\&\\&
  \end{split}
  \end{eqnarray}
\end{subequations}
Here $|0\rangle$ is the vacuum state, $|g\rangle$ is the ground state, and 
$|e_{1,2} \rangle$ are two excited singlet states. The 
$\mathcal{N}_{g/e_2}=(2+2\beta_{\pm}^2)^{-1/2}$ are normalization 
factors and the coefficients $\beta_{\pm}$ are defined as 
\begin{equation} \label{eq:beta}
\beta_\pm=( U \pm \sqrt{16 T  ^2 +U^2})/4 T.
\end{equation} 
The energy eigenvalues corresponding to the eigenstates (\ref{eq:states}) are
\begin{subequations}\label{eq:spectrum}
  \begin{eqnarray}
    &&E_g =2T \beta_- \; , \\
    &&E_{e_1}=U \; ,\\
    &&E_{e_2}=2T \beta_+ \; .
  \end{eqnarray}
\end{subequations}
To simplify notations, we rewrite the external potential part in 
Eq.~\eqref{eq:HubbardH} in the form
\begin{equation} \label{eq:vext}
  \sum_{\sigma} \left( v_1 \hat{n}_{1 \sigma} + v_2 \hat{n}_{2 \sigma} \right) =   \frac{\Delta v}{2} (\hat{n}_{1}  -  \hat{n}_{2}) +C(t) (\hat{n}_{1} +  \hat{n}_{2})
\end{equation} 
where $\hat{n}_{i}=\sum_{\sigma}\hat{n}_{i \sigma}$ is the operator of the 
number of particles on site $i$, $\Delta v= v_1 - v_2$ is the difference of 
on-site potentials, and $C(t)=(v_1(t)+v_2(t))/2$. The last term in 
Eq.~\eqref{eq:vext} corresponds to a spatially uniform potential. This term 
can be trivially gauged away and will be ignored in the following without 
loss of generality. Nontrivial physical effects come only from the external 
potential $\Delta v$ which is coupled to the difference of on-site densities. 
The quantity $\hat d= \hat n_1 - \hat n_2$ can be interpreted as the dipole 
moment of our simplified model of a diatomic system and its expectation value 
$d(t)=\langle\psi(t)|\hat{d}|\psi(t) \rangle$ uniquely determines the on-site densities $n_1(t)$ and $n_2(t)$ if the total number of particles is fixed. In the following, in particular for TDDFT, we will use the dipole moment $d(t)$ as the basic ``density variable''. 

Since the dipole moment operator $\hat d$ is odd under reflection (interchange 
of site indices), it has nonzero matrix elements only between states of 
different parity. In particular, $\hat d$ connects the ground state 
$|g \rangle$ of Eq.~\eqref{eq:GS} only to the first excited state 
$|e_1\rangle$
\begin{equation}\label{eq:dge}
d_{ge}=\langle g| \hat d | e_1 \rangle = \frac{2}{\sqrt{1+ \beta_+ ^2}},
\end{equation} 
while the  matrix element of $\hat d$ between the ground state and the second excited state vanishes, $\langle g | \hat d |e_2\rangle=0$.

Now we are ready to discuss Rabi oscillations in the Hubbard dimer. Let us consider the evolution of the systems from its ground state $|\psi(0)\rangle=|g\rangle$ under the action of a time periodic potential
\begin{equation} \label{eq:laser}
  \Delta v(t)= 2 \E_0 \sin(\omega t).
\end{equation}
The Rabi regime of dynamics occurs when the frequency $\omega$ of a sufficiently weak driving field approaches the frequency $\omega_0$ of the main dipole resonance. In our case this corresponds to the frequency $\omega \sim \omega_0 = E_{e_1}-E_g$ close to the energy difference between ground and first excited states, and the amplitude $ \E_0 \ll  \omega_0/d_{ge}$.

\begin{figure}[ht]
\begin{centering} 
\includegraphics[height=0.34\textwidth,trim = 15mm 00mm 0mm 0mm,clip]{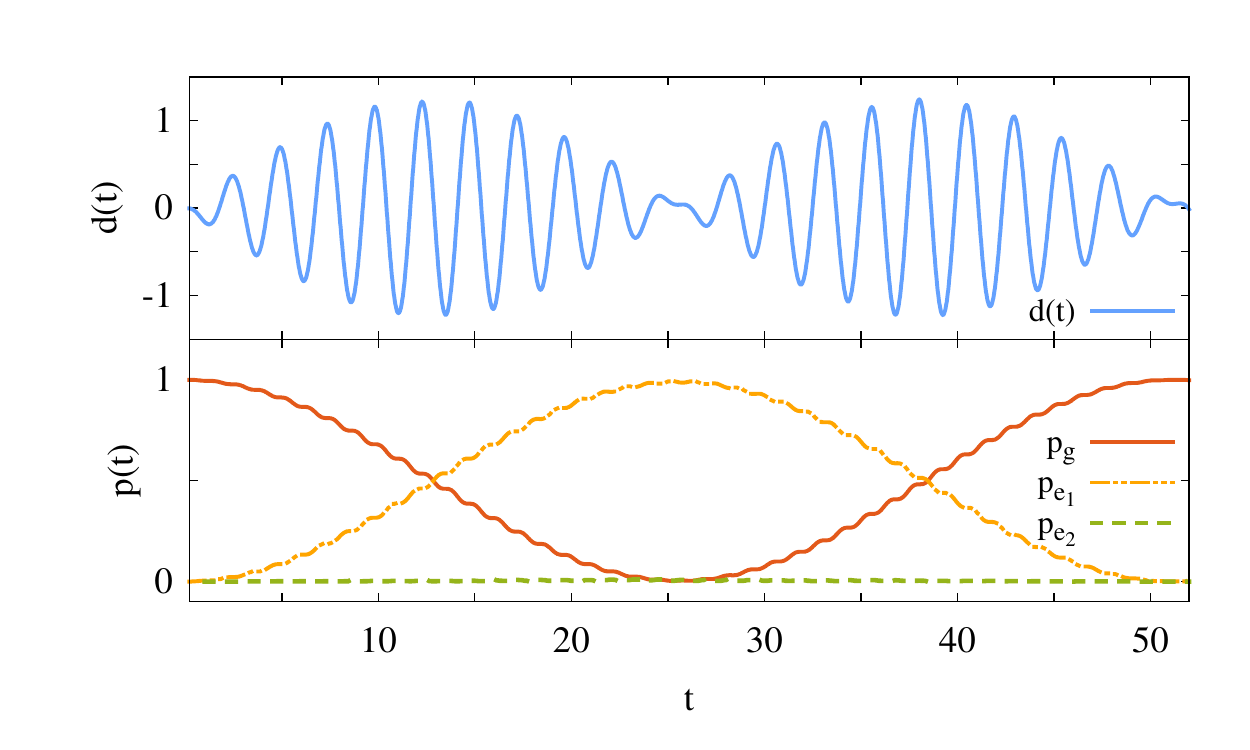}
\caption{(Color online) 
Rabi oscillations for resonant laser $\omega=\omega_0=2.56$~T . 
Upper panel: dipole moment $d(t)$. Lower panel: Population of 
ground state $p_{g}= |\langle g| \psi\rangle|^2$ (solid red),first excited state $p_{e_1}= |\langle e_1| \psi\rangle|^2$ (dotted orange) and second excited state 
$p_{e_2}= |\langle e_2| \psi\rangle|^2$ (dashed green). 
Time is given in units of $1/T$, where T is the hopping parameter.
}
\label{fig:3L-dipole-pop}
\end{centering}
\end{figure}

Fig.~\ref{fig:3L-dipole-pop} shows resonant dynamics of the dipole moment and 
state populations obtained by the numerical propagation of Eq.~\eqref{eq:SE} 
for a moderately strong interaction $U=T=1$, frequency 
$\omega = \omega_0= 2.56$, $d_{ge} = 1.23$, amplitude $\E_0=0.1$ and fixed electron number 
$N=n_1+n_2=2$.
We see that the populations $p_g= |\langle g| \psi\rangle|^2$ and 
$p_{e_1}= |\langle e_1| \psi\rangle|^2$ of the ground and the first exited 
state oscillate between zero and one, while the second excited state stays practically unpopulated, $p_{e_2}= |\langle e_2| \psi\rangle|^2\approx 0$. The dipole moment shows fast oscillations at the driving frequency $\omega$ superimposed with slow oscillations of the envelope at the Rabi frequency $\Omega_R= d_{ge} \E_0$. The maximal value of the dipole moment $|d_{\rm max}|=d_{ge}=1.23$ is reached at 1/4 and 3/4 of the Rabi cycle when the ground and the first excited states have equal populations of 1/2. 

The main characteristic feature of the Rabi regime is a strong variation of 
the state populations. It is this feature which makes the description of Rabi 
oscillations one of the most difficult cases for TDDFT \cite{FHTR11, RB09}. In the rest of this paper we discuss the TDDFT approach to the Rabi dynamics for our simple two-site system.


\section{Time-dependent Kohn-Sham equations for a Hubbard dimer} \label{s:ksHM}

In the present two-electron case the KS system corresponds to two 
non-interacting particles which reproduce the time dependent dipole 
moment $d(t)$ of the interacting system. The KS Hamiltonian has the form of 
Eq.~\eqref{eq:HubbardH} but with no  interaction ($U=0$) and the external 
potential  $\Delta v^s$ is chosen such that the correct time-dependent 
density of the interacting system is reproduced. 

For a singlet state both KS particles occupy the same one-particle KS orbital, 
which is described by two on-site amplitudes $\varphi_1(t)$ and 
$\varphi_2(t)$. Therefore the time-dependent KS equations reduce to a single 
$2\times 2$ one-particle \SE of theform
\begin{subequations}\label{eq:SE-KS}
  \begin{eqnarray}
    i \partial_t \varphi_1= -T \varphi _2+\frac{\Delta v_s}{2} \varphi_1,\\
    i \partial_t \varphi_2=-T \varphi _1 -\frac{\Delta v_s}{2}  \varphi_2 .
  \end{eqnarray}
\end{subequations}
As our dynamics starts from the ground state, Eq.~\eqref{eq:SE-KS} has to be solved with the 
initial condition $\varphi_1(0)=\varphi_2(0)=1/\sqrt{2}$ which corresponds to the noninteracting KS ground state. 
By definition the KS potential $\Delta v_s(t)$ entering Eq.~\eqref{eq:SE-KS} 
produces a prescribed (interacting) dipole moment. In the present case this 
KS potential can be found explicitly as a functional of the density $d(t)$ 
\cite{LiUll2008,FarTok2012},
\begin{equation} \label{eq:vkstddeltan}
  \Delta v_{s} [d] = - \frac{\ddot d +4 T^2 d}{\sqrt{ 4T^2 \left(4 - d^2\right) - \dot{d^2}}}.
\end{equation}
It is important to note that the functional $\Delta v_{s} [d]$ is given by 
Eq.~\eqref{eq:vkstddeltan} only if the system evolves from, and remains 
sufficiently close to, the ground state. 
More precisely, it is shown in Ref.~\cite{FarTok2012} that the 
functional form of Eq.~\eqref{eq:vkstddeltan} is valid as long as the 
condition $|\arg(\varphi_1)-\arg(\varphi_2)|< \pi/2$ is satisfied during the 
course of the evolution. If the opposite inequality holds, the overall 
sign on the right hand side of Eq.~\eqref{eq:vkstddeltan} has to be changed 
from $-$ to $+$. Moreover, the sign changes every time the line  
$|\arg(\varphi_1)-\arg(\varphi_2)|=\pi/2$ is crossed. In terms of the dipole 
moment, crossing this line corresponds to a vanishing expression under the 
square root in Eq.~\eqref{eq:vkstddeltan} 
\footnote{Since the whole right hand side in this equation
is proportional to the time derivative of the denominator, turning the square root into zero does not lead to a divergence because the numerator vanishes at the same time. Also, for any physical evolution the expression under the square root always stays nonnegative as $|\dot d|$ for a lattice is properly bounded from above ~\protect\cite{V08,TokatlyL2011,FarTok2012}. }.
The above behavior can be viewed as a manifestation of the initial state and history dependence in TDDFT \cite{MaitraTDDFT2011,FarTok2012}. 

The exact KS potential can be calculated by inserting into 
Eq.~\eqref{eq:vkstddeltan} the exact dipole moment $d(t)$ obtained from a 
numerical solution of the many-body \SE \eqref{eq:SE}. In order to get the 
Hartree-exchange-correlation (\hxc) potential we subtract the physical 
external potential $\Delta v$ from the KS potential,
\begin{equation} \label{eq:ex_Dvhxc}
\Delta v_{\hxc}  = \Delta v_{s} - \Delta v.
\end{equation}
The time dependence of the exact $\Delta v_{\hxc}$ which corresponds to the 
dipole moment $d(t)$ presented in Fig.~1 (i.~e., to the regime of resonant 
Rabi oscillations, described in Sec.~II) is shown in the top panel of 
Fig.~\ref{fig:non_adiab}.

In practice the exact Hxc functional is unknown and one has to rely on 
approximations. The simplest and the most common approximation in TDDFT is 
based on the adiabatic assumption for xc effects. Below we present and test 
the adiabatic approximation for our model system. 


\subsection{Adiabatically exact functional}

To construct the adiabatic approximation for the Hubbard dimer we first find 
the exact ground-state \hxc functional by the Levy-Lieb constrained search, 
i.e., we perform an exhaustive search over the space of all allowed 
two-particle wave functions $\Psi$ that yield a given dipole moment $d$ to 
find the Hohenberg-Kohn energy functional $F_{\sss HK}[d]$,  
\begin{eqnarray} \label{eq:cs}
   F_{\sss HK}[d]= \min_{\Psi \rightarrow d}\langle \Psi|\hat{T} + \hat{U}|\Psi\rangle,
\end{eqnarray}
where  $\hat{T}$ and $\hat{U}$ are operators of the kinetic energy and the interaction energy, i.~e., the first and the second terms in the Hamiltonian \eqref{eq:HubbardH},  respectively. 

The exact ground state \hxc potential is given by the derivative of the Hxc energy with respect to the dipole moment,
\begin{equation} \label{eq:vhxc}  
\Delta v_{\sss \hxc}^{gs}[d] =2~ \dfrac{\partial}{\partial{d}} \big( F_{\sss HK}[d]-T^s[d]\big)
\end{equation}
where $T^s[d]$ is the kinetic energy functional that is defined by Eq.~\eqref{eq:cs} with $U=0$. More details on this construction can be found in the Appendix.

In the {\it adiabatically-exact} approximation the exact ground-state \hxc 
potential of Eq.~(\ref{eq:vhxc}) is used in the time-dependent KS equations, 
i.e., the Hxc potential at time $t$ is calculated by inserting the 
instantaneous value of $d(t)$ into the ground-state functional 
\begin{equation}
\Delta v_{\sss \hxc}^{ad}[d](t)= \Delta v_{\sss \hxc}^{gs}[d(t)].
\end{equation}
 
We note that the adiabatically-exact \hxc potential has also been found 
numerically for real-space one-dimensional two-electron systems in 
Refs.~\cite{TGK08,FERM13, EFRM12} using the iterative procedure introduced in 
Ref.~\cite{TGK08}. A big advantage of the present simple model is that we know 
explicitly the functional dependence of the exact ground-state \hxc potential 
(see Appendix), i.~e., we do not need any a priori knowledge of the 
time-dependent density.

To test the performance of the adiabatically-exact functional in the regime 
of Rabi oscillations we propagate self-consistently the KS equations with 
$v_{\hxc}^{ad}[d](t)$ for the same parameters as in Sec.~\ref{s:HM}.
\begin{figure}[ht]
\begin{centering}
\includegraphics[height=0.34\textwidth,trim = 15mm 00mm 0mm 0mm,clip]{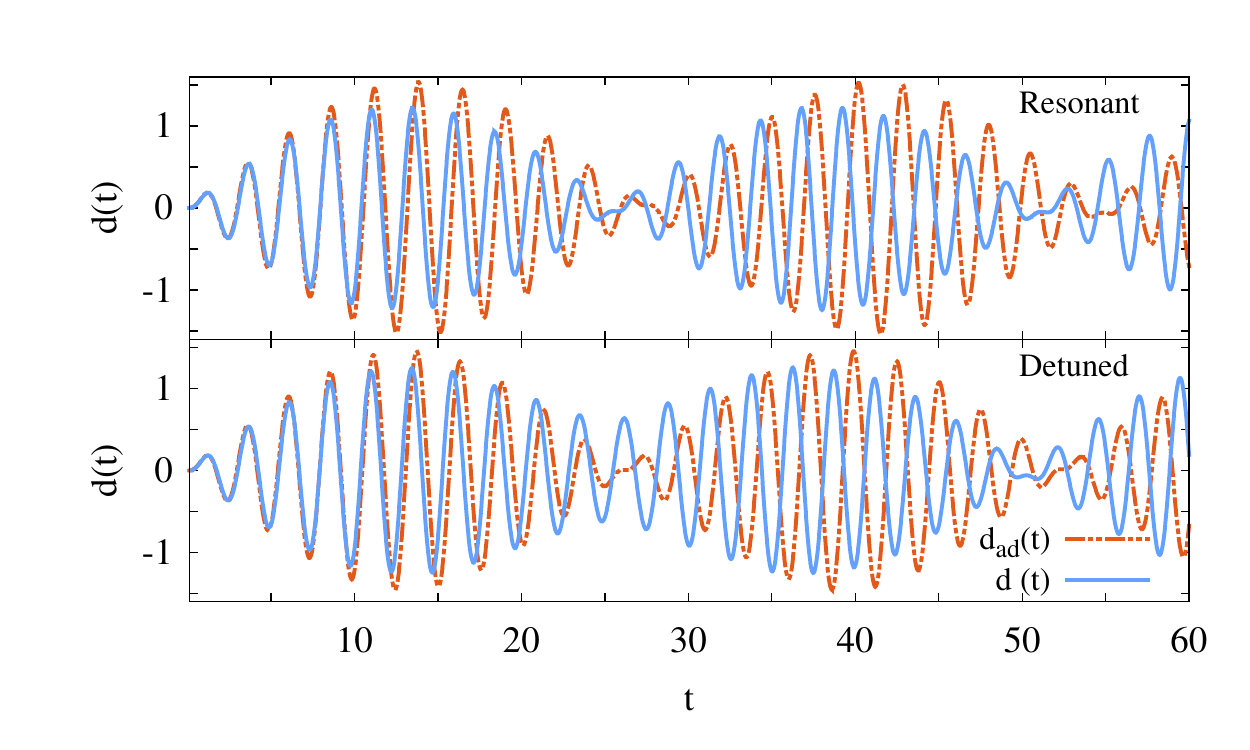}
\caption{(Color online) Upper panel : $d(t)$ (solid blue) in the presence of a laser of frequency $\omega=\omega_0=2.56$~T, compared to $d_{\rm ad}(t)$ (dotted red) propagated using the exact ground state functional $\Delta v^{gs}_{\hxc}[d]$ in the presence of a laser resonant with the  adiabatically-exact linear response frequency $\omega_{\rm ad}^{LR}=2.60$~T. Lower panel: $d(t)$ for slightly detuned laser $\omega=\omega_0+0.03$ (solid blue) compared to $d_{\rm ad}(t)$ using $\omega=\omega_{\rm ad}^{LR}+0.03$ (dotted red). Time is given in units of the inverse of the hopping parameter $T$.
}
\label{fig:dd-resonant-det}
\end{centering}
\end{figure}

The results of the propagation confirm a general conclusion of 
Ref.~\cite{FHTR11} about the presence of an artificial dynamical detuning in 
the description of Rabi oscillations using adiabatic functionals. 
In Fig.~\ref{fig:dd-resonant-det}  we 
compare the evolution of the exact dipole moment $d(t)$ (blue) with 
the dipole moment $d_{\rm ad}(t)$ (red) obtained from KS equations with the 
adiabatically-exact Hxc potential for resonant (upper panel) and slightly detuned (lower panel) applied lasers.  
The upper panel shows 
the dynamics at resonant conditions when the frequency $\omega$ of the 
driving field is equal to the frequency $\omega_{\rm res}$ of the main dipole 
resonance. In the exact interacting system this frequency is obviously 
$\omega_{\rm res}=\omega_0$, while in the approximate TDDFT it is 
approximation-dependent, and should be determined consistently as the 
frequency $\omega_{\rm res}=\omega_{LR}$ of the corresponding linear response 
resonance. At first sight, the function  $d_{\rm ad}(t)$ resulting from 
TDDFT with the adiabatic Hxc potential (upper panel in red on 
Fig.~\ref{fig:dd-resonant-det}) looks qualitatively similar to the exact $d(t)$ 
(upper panel in blue on Fig.~\ref{fig:dd-resonant-det}). However, there is a deep 
difference in the underlying microscopic dynamics. The physical system returns 
to its initial state after two periods of the dipole moment's envelope, which 
corresponds to the Rabi period of $T_{\rm R} = \frac{2 \pi}{{\cal E}_0 d_{ge}}= 51.10$. In contrast, the 
microscopic period of the KS system with the adiabatic Hxc potential coincides 
with that of the dipole moment, which is the characteristic feature of 
detuned Rabi oscillations. In fact, the KS Rabi dynamics is always internally 
detuned by the presence of the adiabatic potential $\Delta v_{\hxc}^{gs}[d(t)]$ 
which depends on the instantaneous density \cite{FHTR11}.  While this 
important difference is hidden in the case of resonant dynamics of the dipole 
moment, it is revealed immediately when the driving frequency is a bit 
shifted (detuned) from the exact resonance. The dipole moments $d(t)$ and 
$d_{\rm ad}(t)$ for a slightly detuned driving field with 
$\omega=\omega_{\rm res}+0.03$ are presented in the lower panel on Fig.~\ref{fig:dd-resonant-det}. The 
exact dipole moment $d(t)$ develops a ``neck'' at $t\sim T_{\rm R}$ showing 
that the actual physical period is indeed $T_{\rm R}\approx 50$. On the other 
hand, the function $d_{\rm ad}(t)$ is practically unaffected by the external 
detuning because the KS system, being already strongly detuned internally, is 
insensitive to small external variations of the driving frequency.
This qualitative failure of the adiabatic approximation clearly demonstrates 
the important role of xc memory effects in the correct description of Rabi 
oscillations.

\begin{figure}[ht]
\includegraphics[height=0.34\textwidth,trim = 15mm 00mm 0mm 0mm,clip]{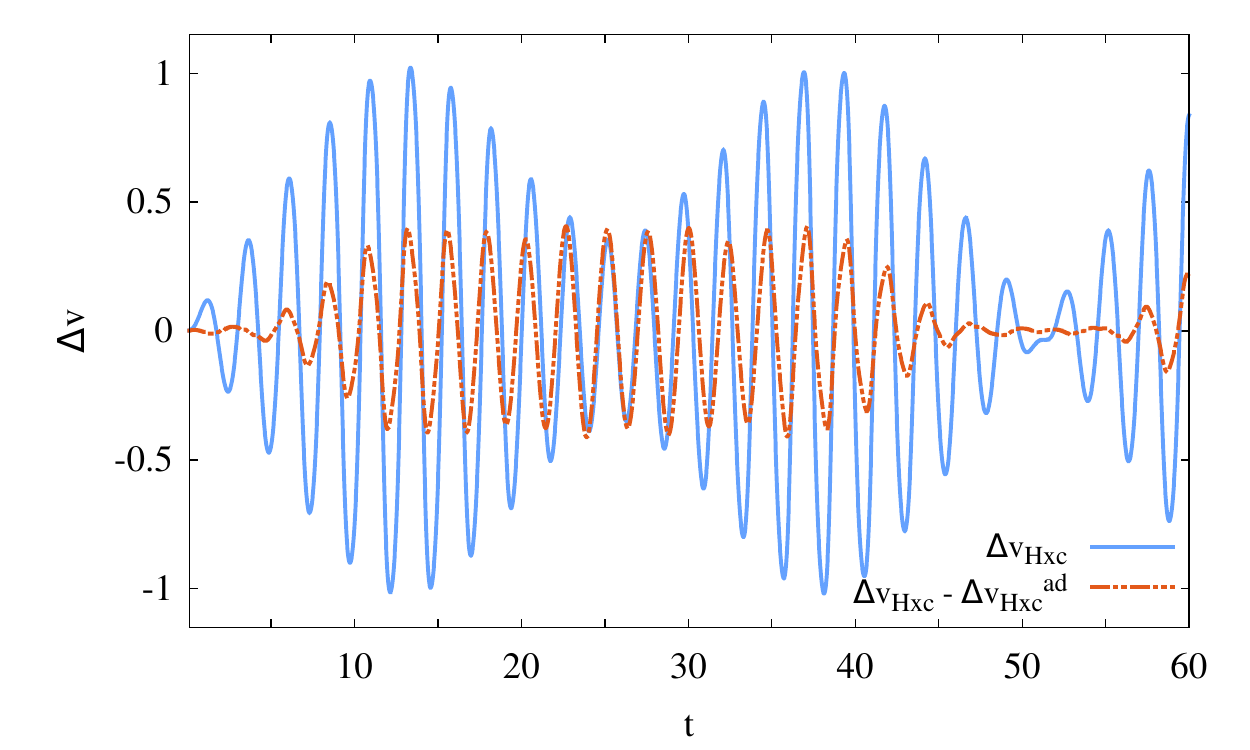}
\caption{(Color online) Time-dependent Hxc potential $\Delta v_{\hxc}(t)$ (in units of the hopping parameter $T$) (solid blue) and its non-adiabatic  
contribution defined as $\Delta v_{\hxc}(t)-\Delta v_{\hxc}^{gs}[d(t)]$ (in units of the hopping parameter $T$) (dotted red). Time is given in units of $1/T$.}
\label{fig:non_adiab}
\end{figure}

To further quantify non-adiabatic effects in the Rabi regime we extract a non-adiabatic contribution to the total \hxc potential. Namely, we subtract the adiabatic potential $\Delta v_{\hxc}^{gs}[d(t)]$ evaluated at the exact dipole moment $d(t)$ from the exact $\Delta v_{\hxc}(t)$ defined by Eqs.~\eqref{eq:ex_Dvhxc} and \eqref{eq:vkstddeltan}. In \fig \ref{fig:non_adiab} we present the non-adiabatic part of \hxc potential together with the exact $\Delta v_{\hxc}(t)$. 
The non-adiabatic contribution to $\Delta v_{\hxc}(t)$ turns out to be more than double the amplitude of the external potential and in fact as large as the \hxc potential itself during a significant part of the Rabi-cycle.
Not surprisingly, the non-adiabatic effects are small at the beginning and at the end of the Rabi cycle when the ground-state population is dominant and the system is close to the linear response regime. But they grow fast when the system is driven away from the ground-state and they remain large for a large part of the Rabi-cycle.  It is interesting to notice that, centered around $T_R/2$, there is a long period of time during which the amplitude of the adiabatic effects remains almost constant (see Fig.~\ref{fig:non_adiab}).

Apparently a better approximation for the xc potential is needed to capture non-adiabatic effects relevant to describe Rabi oscillations. In the next section we propose an explicit non-adiabatic density functional based on a two-level description of the interacting system.


\section{Time-dependent xc potential in the two-level approximation}
\label{s:exhxcF}

In general, the Hxc functional $\Delta v_{\hxc}[d]$ can be found via Eq.~\eqref{eq:ex_Dvhxc} if we know the external potential as a functional of $d(t)$.  The presence of interactions makes the problem of finding the functional $\Delta v[d]$ highly nontrivial even in our simple model. Fortunately in some cases like the Rabi oscillations the problem is simplified dramatically because the behavior of the system is close to that of an effective two-level system. 

Let us assume that the second excited state $|e_2\rangle$ is not participating 
in the dynamics. We write the many-body \SE \eqref{eq:SE} in the two level 
approximation as 
\begin{subequations} \label{eq:SE2l}
  \begin{eqnarray}
    i \partial_t \psi_g(t)= E_g \psi_g (t) + d_{ge} \frac{\Delta v(t)}{2} \psi_{e}(t) \; ,\\
    i \partial_t \psi_{e}(t)= E_{e_1} \psi_{e}(t)+ d_{ge} \frac{\Delta v(t)}{2} \psi_{g}(t) \; ,
  \end{eqnarray}
\end{subequations}
where $\psi_g(t)=\langle g|\psi(t)\rangle$, and 
$\psi_e(t) = \langle e_1|\psi(t)\rangle$ are the projections of the 
time-dependent wave function onto the ground and first excited state, 
respectively. 

By rotating the basis 
we can represent Eq.~\eqref{eq:SE2l} in the form of a \SE for one particle on an effective ``two-site lattice''. In other words, Eq.~\eqref{eq:SE2l} is unitarily equivalent to Eq.~\eqref{eq:SE-KS} with hopping constant $\omega_0/2$ and external potential given by $d_{ge}\Delta v/2$. Using this mapping and the KS potential $\Delta v_s$ of Eq.~\eqref{eq:vkstddeltan} we can immediately write the external potential $\Delta v$ of the interacting system as a functional of the dipole moment $d$
\begin{equation} \label{eq:vextdeltan}
  \Delta v^{2L}[d] =\frac{(-1)^p}{d_{ge}} \left(\frac{\ddot d + \omega_0 ^2 d}{\sqrt{\omega_0 ^2 (d_{ge}^2-d^2) -{\dot d}^2  }}\right),
\end{equation}
where the integer $p$ counts how many times the square root turns into zero during the evolution. The factor $(-1)^p$ accounts for the sign changes explained after Eq.~\eqref{eq:vkstddeltan}. 

In order to find the Hxc potential $\Delta v_{\hxc}[d]$ as a functional of the dipole moment  we substitute the external potential $\Delta v $ of Eq.~\eqref{eq:vextdeltan} and the KS potential of Eq.~\eqref{eq:vkstddeltan} into Eq.~\eqref{eq:ex_Dvhxc}:
\begin{eqnarray}\label{eq:vhxc_2l}
  \Delta v_{\hxc} ^{2L} [d] = &-& \frac{\ddot d +4 T^2 d}{\sqrt{ 4T^2 \left(4 - d^2\right) - \dot{d^2}}}\nn\\ 
  &-&\frac{(-1)^p}{d_{ge}} \left(\frac{\ddot d + \omega_0 ^2 d}{\sqrt{\omega_0^2 (d_{ge}^2-d^2) -{\dot d}^2  }}\right)
\end{eqnarray}
This expression is one of the main results of the present paper. It provides us with an explicit fully non-adiabatic density functional which, by construction, should correctly describe the Rabi oscillations. It is worth emphasizing that the functional Eq.~\eqref{eq:vhxc_2l} contains history dependence via the integer $p$ in the second term.


One can easily check that the non-linear functional $\Delta v_{\hxc}^{2L}$ of Eq.~\eqref{eq:vhxc_2l} produces the exact dynamic xc kernel in the linear response regime. The formally exact \hxc functional in the linear response can be written as follows 
\begin{equation}\label{eq:vhxcLR1}
  \Delta v_{\hxc} ^{LR}[d](\omega)=\big[\chi_s^{-1}(\omega)-\chi^{-1}(\omega) \big] d
  = f_{xc}(\omega )d,
\end{equation}
where $\chi_s(\omega)$ and $\chi(\omega)$ are the density response functions 
for the KS and the interactiving system, respectively. By definition the term 
in parentheses is the exact exchange-correlation kernel $f_{xc}(\omega)$. Since the eigenfunctions for the Hubbard dimer are known, Eqs.~\eqref{eq:states} and \eqref{eq:spectrum}, we can write the exact response functions $\chi(\omega)$ in the Lehmann representation \cite{VignaleBook}, and substitute it into Eq.~\eqref{eq:vhxcLR1}. The result takes the following form
\begin{equation} 
\label{eq:vhxcLR2}
  \Delta v_{\hxc}^{LR}[d](\omega) =\left( T - \frac{\omega_0}{2 d_{ge}^2}  + \big(\frac{1}{4 T} - \frac{1}{2 \omega_0 d_{ge}^2}\big) \omega^2 \right) d .
\end{equation}
It is now straightforward to see that this equation is identical to the linearized 
version of the approximate functional $\Delta v_{\hxc} ^{2L} [d]$ defined by 
Eq.~\eqref{eq:vhxc_2l} with $p=0$. In other words our approximation becomes 
exact in the linear regime. This nice property is not accidental because the 
functional of Eq.~\eqref{eq:vhxc_2l} is based on the two-level approximation. 
In the linear response regime, the symmetric Hubbard 
dimer becomes an effective two-level system because the dipole transition 
matrix element between the ground state $|g\rangle$ and the second excited 
state $|e_2\rangle$ vanishes.

A subtle property of the non-adiabatic functional Eq.~\eqref{eq:vhxc_2l} is the dependence on the second time derivative $\ddot d$ of the dipole moment. The presence of $\ddot d$ does not mean that the xc potential assumes a dependence on the future. In general the existence theorem  for TDDFT on a lattice \cite{FarTok2012} requires the second time derivative of the density to be continuous. Therefore $\ddot d$ can be calculated as a left limit for any time greater than the initial time, $t>0$. At $t=0$ the value of $\ddot d(0)$ is determined by the initial value of the external potential as follows
\begin{equation}
  \ddot d(0)= -d_{de}\Delta v(0) \sqrt{\omega_0 ^2 \left(d_{ge}^2-d^2(0)\right) -{\dot d^2 (0)} }  -\omega _0 ^ 2 d(0),
\end{equation}
where the $d(0)$ and $\dot d(0)$ are fixed by the KS-initial state
\begin{subequations}
  \begin{eqnarray}
    d(0)&=& 2(|\varphi_1(0)|^2-|\varphi_2(0)|^2),\\
    \dot d (0)&=&-4 T {\rm Im}[\varphi_1^*(0)\varphi_2 (0)].
  \end{eqnarray}
\end{subequations}

After this preliminaries we can plug the Hxc potential Eq.~\eqref{eq:vhxc_2l} 
into the KS equations and propagate them self-consistently to test the 
performance of our non-adiabatic approximation. It is, however, clear that 
the functional $\Delta v_{\hxc} ^{2L} [d]$, by construction, should exactly 
reproduce the results of the two-level approximation to the full interacting 
problem. Therefore TDDFT with the Hxc potential of Eq.~\eqref{eq:vhxc_2l} is 
as accurate as the two-level approximation itself.
\begin{figure}[ht]
\includegraphics[height=0.34\textwidth,trim = 15mm 00mm 0mm 0mm,clip]{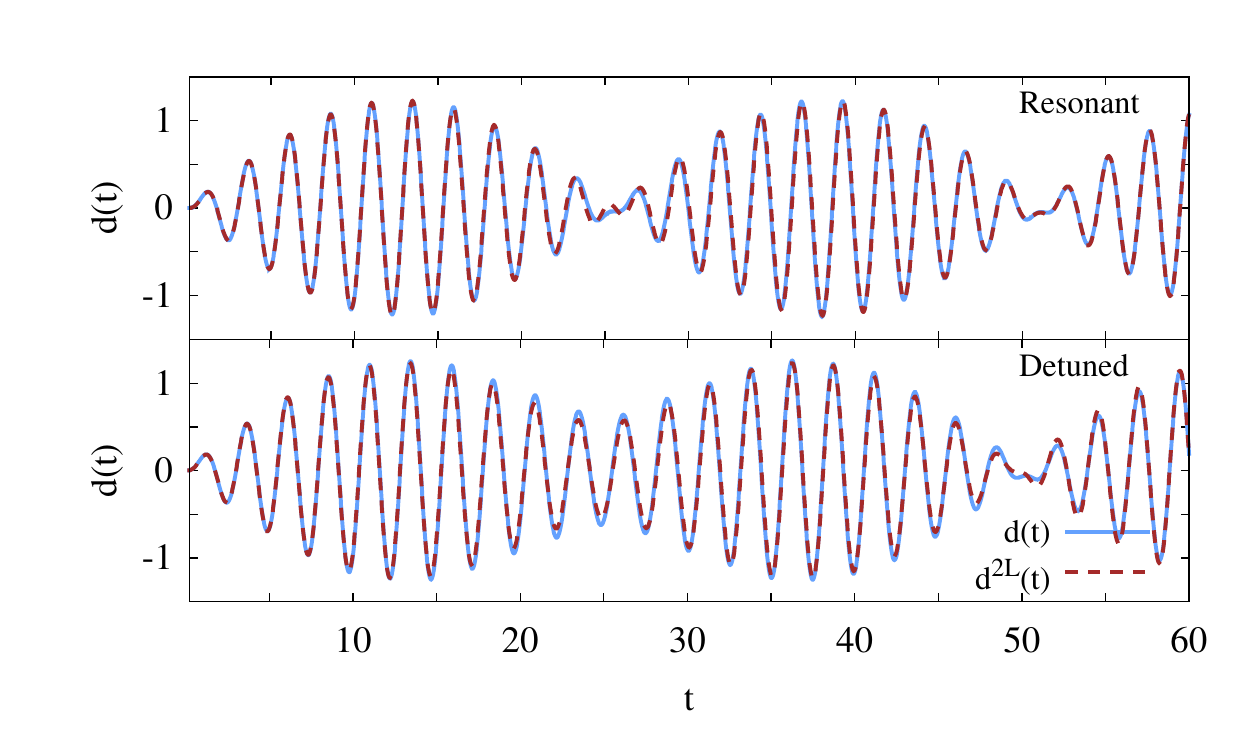}
\caption{(Color online) Upper panel: $d(t)$ (solid blue) for resonant laser frequency $\omega=\omega_0=2.56$~T compared to two-level approximation $d^{2L}(t)$ (dashed brown) using $\Delta v_{\hxc}^{2L}$, Eq.~\eqref{eq:vhxc_2l}, and same laser frequency.  Lower panel: $d(t)$ (solid blue) and $d^{2L}(t)$ (dashed red) for slightly detuned laser $\omega=\omega_0+0.03$~T (same detuning as lower panel on Fig.~\ref{fig:dd-resonant-det}).
 Time is given in units of $1/T$.}
\label{fig:2lvs3l}
\end{figure}

In \fig \ref{fig:2lvs3l} we compare the exact resonant Rabi dynamics of the dipole moment with the one obtained in the two-level approximation or, alternatively, by solving self-consistently the KS equations with the potential $\Delta v_{\hxc} ^{2L} [d]$. The exact and approximate dipole moments are practically on top of each other. The non-adiabatic functional of Eq.~\eqref{eq:vhxc_2l} excellently reproduces Rabi oscillations for a resonant excitation. Apparently it also works perfectly for detuned Rabi dynamics provided the detuning is not too large. Another nice property of this approximation is that it becomes essentially exact for a sufficiently weak non-resonant driving potential which corresponds to the linear response regime.

\section{Conclusion}
\label{s:conclus}

We use a Hubbard dimer to analyze, both qualitatively and quantitatively, the non-adiabatic features present in the TDDFT functional. For this model system the exact Kohn-Sham potential is analytic and moreover, the exact ground state functional can be found by Levy-Lieb constrained search. The later is propagated self-consistently to study the performance of the adiabatic approximation. We show that non-adiabaticity is crucial to properly capture the physics of resonant and nearly detuned Rabi oscillations.  The observed non-adiabatic features grow as the population of the excited state is rising to its maximum, becoming even larger than the external potential. Lack of these features in adiabatic functionals causes them to fail to describe Rabi dynamics, missing both frequency and amplitude of the physical dipole moment.
Taking advantage of the fact that under the action of a resonant laser the system behaves as an effective two-level one, we derive an explicit non-adiabatic functional that accurately reproduces resonant and slightly detuned Rabi oscillations.
This fully non-adiabatic functional incorporates explicitely the initial-state dependence and becomes exact in the linear response regime. 
\\
The present work was focussed on the TDDFT description of Rabi oscillations 
in a minimal model system. While the construction of the quasi-exact TDDFT for 
this minimalistic system was non-trivial, the non-adiabatic part of the Hxc 
potential has been found to have a relatively simple structure (see Fig.~\ref{fig:non_adiab}): while non-adiabaticity is small close to the beginning and the end of the Rabi cycle, its amplitude is signifcant but almost constant throughout a large middle part of the Rabi cycle. This might be a useful observation when aiming to construct non-adiabatic TDDFT functionals applicable to realistic systems. 

We acknowledge financial support from the European Research Council Advanced Grant DYNamo (ERC-2010-AdG-267374), Spanish Grant (FIS2010-21282-C02-01), Grupos Consolidados UPV/EHU del Gobierno Vasco (IT578-13),  Ikerbasque and the  European Commission projects CRONOS (Grant number 280879-2 CRONOS CP-FP7). 

\appendix*

\section{Exact Hohenberg-Kohn functional of the two-site Hubbard model 
by constrained search}
\label{s:append}

In this Appendix we briefly describe how one can construct the exact 
Hohenberg-Kohn functional (\ref{eq:cs}) for two electrons in the two-site 
Hubbard model by carrying out the constrained search as suggested by Levy 
\cite{Levy:79,Levy82} and Lieb \cite{Lieb:83}. 

The Hilbert space for two fermions on two sites is of dimension six and  
separates into a singlet and a triplet sector of dimension three each. Since 
for any value of $d$, the ground state of $\hat{H}_0 = \hat{T} + \hat{U}$ is 
a singlet, we may restrict the search in Eq.~(\ref{eq:cs}) to singlet 
wavefunctions only. As a basis for the singlet sector we use the 
eigenstates of Eq.~(\ref{eq:states}). Then the most general singlet state 
may be written as 
\be
| \Psi \rangle = A_1 |g \rangle + A_2 |e_1 \rangle + A_3 |e_2 \rangle,
\label{singlet_general}
\ee
where we can, without loss of generality, choose the coefficients $A_i$ to be 
real. In the chosen basis, the expectation value of 
$\hat{H}_0=\hat{T}+\hat{U}$ takes the simple form
\be
\langle \Psi | \hat{H}_0 | \Psi \rangle = 
\langle \Psi | \hat{T} + \hat{U} |\Psi \rangle = 
E_g A_1^2 + E_{e_1} A_2^2 + E_{e_2} A_3^2 
\ee
where the eigenvalues $E_g$, $E_{e_1}$, and $E_{e_2}$ of 
the basis functions are given by Eq.~(\ref{eq:spectrum}). 

\begin{figure}
\begin{center}
\includegraphics[height=0.34\textwidth,trim = 15mm 00mm 0mm 0mm,clip]{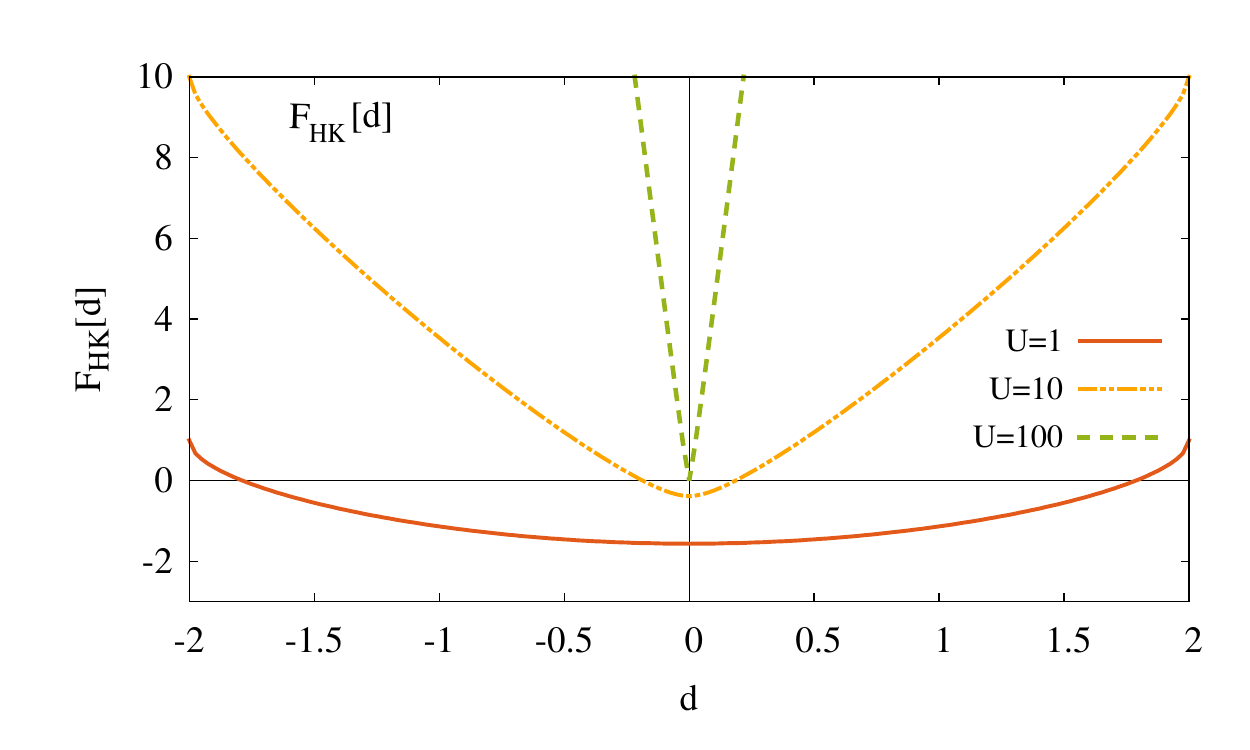}
\caption{ Exact Hohenberg Kohn density functional $F_{\rm HK}[d]$ (in units of the hopping parameter $T$) for different Hubbard strength's $U$.} 
\label{fig:FHK}
\end{center}
\end{figure}

The expansion coefficients $A_i$ in Eq.~(\ref{singlet_general}) are not 
independent. The normalization condition of the wavefunction $|\Psi \rangle$ 
leads to
\be
A_1^2 + A_2^2 + A_3^2 = 1 \;.
\label{normalize}
\ee
In the constrained search we also have to make sure 
that we are only searching over wavefunctions which yield a given ``density'' 
$d$. This gives a second condition on the coefficients which reads
\be
d = \langle \Psi | \hat{n_1}-\hat{n}_2 | \Psi \rangle = 
4 \frac{A_1 + \beta_{+} A_3}{\sqrt{1 + \beta_{+}^2}} A_2
\label{density_cond}
\ee
where $\beta_{+}$ is given by Eq.~(\ref{eq:beta}).
We can use Eqs.~(\ref{normalize}) and (\ref{density_cond}) to eliminate two 
of the coefficients, say $A_2$ and $A_3$, in the constrained search which then 
becomes a minimization in a single variable, i.e., 
\be
F_{\rm HK}(d) = \min_{A_1} \; \langle \Psi(A_1,d) | \hat{T} + \hat{U} | 
\Psi(A_1,d) \rangle \;.
\ee
In general, this minimization has to be carried out numerically. In 
Fig.~\ref{fig:FHK} we show $F^{\rm HK}$ as function of the dipole moment for 
various values of $U$. We note that $F^{\rm HK}(d)$ is always minimal at 
$d=0$. For large values of $U$ the slope of $F^{\rm HK}(d)$ changes rapidly 
as one crosses from negative to positive values of $d$. For vanishing 
interaction $U=0$, the minimization can be carried out fully analytically. The 
resulting functional, the non-interacting kinetic energy, reads
\be
T_s(d) = \min_{\Psi \to d} \langle \Psi | \hat{T} |\Psi \rangle = 
2 T \left( 1 - 2 \sqrt{1 - \frac{d^2}{8}}\right) \; .
\ee
The Hartree-exchange-correlation energy then is given by
\be
E_{\rm \hxc}(d) = F_{\rm HK}(d) - T_s(d)
\ee
and the corresponding \hxc potential can be easily obtained by differentiation. 


\bibliography{2site-Hubbard_v10}

\begin{thebibliography}{33}%
\makeatletter
\providecommand \@ifxundefined [1]{%
 \@ifx{#1\undefined}
}%
\providecommand \@ifnum [1]{%
 \ifnum #1\expandafter \@firstoftwo
 \else \expandafter \@secondoftwo
 \fi
}%
\providecommand \@ifx [1]{%
 \ifx #1\expandafter \@firstoftwo
 \else \expandafter \@secondoftwo
 \fi
}%
\providecommand \natexlab [1]{#1}%
\providecommand \enquote  [1]{``#1''}%
\providecommand \bibnamefont  [1]{#1}%
\providecommand \bibfnamefont [1]{#1}%
\providecommand \citenamefont [1]{#1}%
\providecommand \href@noop [0]{\@secondoftwo}%
\providecommand \href [0]{\begingroup \@sanitize@url \@href}%
\providecommand \@href[1]{\@@startlink{#1}\@@href}%
\providecommand \@@href[1]{\endgroup#1\@@endlink}%
\providecommand \@sanitize@url [0]{\catcode `\\12\catcode `\$12\catcode
  `\&12\catcode `\#12\catcode `\^12\catcode `\_12\catcode `\%12\relax}%
\providecommand \@@startlink[1]{}%
\providecommand \@@endlink[0]{}%
\providecommand \url  [0]{\begingroup\@sanitize@url \@url }%
\providecommand \@url [1]{\endgroup\@href {#1}{\urlprefix }}%
\providecommand \urlprefix  [0]{URL }%
\providecommand \Eprint [0]{\href }%
\providecommand \doibase [0]{http://dx.doi.org/}%
\providecommand \selectlanguage [0]{\@gobble}%
\providecommand \bibinfo  [0]{\@secondoftwo}%
\providecommand \bibfield  [0]{\@secondoftwo}%
\providecommand \translation [1]{[#1]}%
\providecommand \BibitemOpen [0]{}%
\providecommand \bibitemStop [0]{}%
\providecommand \bibitemNoStop [0]{.\EOS\space}%
\providecommand \EOS [0]{\spacefactor3000\relax}%
\providecommand \BibitemShut  [1]{\csname bibitem#1\endcsname}%
\let\auto@bib@innerbib\@empty
\bibitem [{\citenamefont {Marques}\ \emph {et~al.}(2012)\citenamefont
  {Marques}, \citenamefont {Maitra}, \citenamefont {Nogueira}, \citenamefont
  {Gross},\ and\ \citenamefont {Rubio}}]{TDDFTBook2012}%
  \BibitemOpen
  \bibfield  {author} {\bibinfo {author} {\bibfnamefont {M.}~\bibnamefont
  {Marques}}, \bibinfo {author} {\bibfnamefont {N.~T.}\ \bibnamefont {Maitra}},
  \bibinfo {author} {\bibfnamefont {F.}~\bibnamefont {Nogueira}}, \bibinfo
  {author} {\bibfnamefont {E.}~\bibnamefont {Gross}}, \ and\ \bibinfo {author}
  {\bibfnamefont {A.}~\bibnamefont {Rubio}},\ }\href@noop {} {\emph {\bibinfo
  {title} {Fundamentals of time-dependent density functional theory}}},\
  \bibinfo {series} {Lecture Notes in Physics}, Vol.\ \bibinfo {volume} {837}\
  (\bibinfo  {publisher} {Springer},\ \bibinfo {address} {Heidelberg},\
  \bibinfo {year} {2012})\BibitemShut {NoStop}%
\bibitem [{\citenamefont {Casida}(2009)}]{Casida09}%
  \BibitemOpen
  \bibfield  {author} {\bibinfo {author} {\bibfnamefont {M.~E.}\ \bibnamefont
  {Casida}},\ }\href {\doibase
  http://dx.doi.org/10.1016/j.theochem.2009.08.018} {\bibfield  {journal}
  {\bibinfo  {journal} {Journal of Molecular Structure: THEOCHEM}\ }\textbf
  {\bibinfo {volume} {914}},\ \bibinfo {pages} {3 } (\bibinfo {year}
  {2009})}\BibitemShut {NoStop}%
\bibitem [{\citenamefont {Livshits}\ and\ \citenamefont {Baer}(2007)}]{LB07}%
  \BibitemOpen
  \bibfield  {author} {\bibinfo {author} {\bibfnamefont {E.}~\bibnamefont
  {Livshits}}\ and\ \bibinfo {author} {\bibfnamefont {R.}~\bibnamefont
  {Baer}},\ }\href {\doibase 10.1039/B617919C} {\bibfield  {journal} {\bibinfo
  {journal} {Phys. Chem. Chem. Phys.}\ }\textbf {\bibinfo {volume} {9}},\
  \bibinfo {pages} {2932} (\bibinfo {year} {2007})}\BibitemShut {NoStop}%
\bibitem [{\citenamefont {Maitra}(2005)}]{M05}%
  \BibitemOpen
  \bibfield  {author} {\bibinfo {author} {\bibfnamefont {N.~T.}\ \bibnamefont
  {Maitra}},\ }\href {\doibase 10.1063/1.1924599} {\bibfield  {journal}
  {\bibinfo  {journal} {J. Chem. Phys.}\ }\textbf {\bibinfo {volume} {122}},\
  \bibinfo {eid} {234104} (\bibinfo {year} {2005})}\BibitemShut {NoStop}%
\bibitem [{\citenamefont {Gritsenko}\ and\ \citenamefont
  {Baerends}(2004)}]{GB04}%
  \BibitemOpen
  \bibfield  {author} {\bibinfo {author} {\bibfnamefont {O.}~\bibnamefont
  {Gritsenko}}\ and\ \bibinfo {author} {\bibfnamefont {E.~J.}\ \bibnamefont
  {Baerends}},\ }\href {\doibase 10.1063/1.1759320} {\bibfield  {journal}
  {\bibinfo  {journal} {J. Chem. Phys.}\ }\textbf {\bibinfo {volume} {121}},\
  \bibinfo {pages} {655} (\bibinfo {year} {2004})}\BibitemShut {NoStop}%
\bibitem [{\citenamefont {Tozer}(2003)}]{T03}%
  \BibitemOpen
  \bibfield  {author} {\bibinfo {author} {\bibfnamefont {D.~J.}\ \bibnamefont
  {Tozer}},\ }\href {\doibase 10.1063/1.1633756} {\bibfield  {journal}
  {\bibinfo  {journal} {J. Chem. Phys.}\ }\textbf {\bibinfo {volume} {119}},\
  \bibinfo {pages} {12697} (\bibinfo {year} {2003})}\BibitemShut {NoStop}%
\bibitem [{\citenamefont {Maitra}\ \emph {et~al.}(2004)\citenamefont {Maitra},
  \citenamefont {Zhang}, \citenamefont {Cave},\ and\ \citenamefont
  {Burke}}]{MZCB2004}%
  \BibitemOpen
  \bibfield  {author} {\bibinfo {author} {\bibfnamefont {N.~T.}\ \bibnamefont
  {Maitra}}, \bibinfo {author} {\bibfnamefont {F.}~\bibnamefont {Zhang}},
  \bibinfo {author} {\bibfnamefont {R.~J.}\ \bibnamefont {Cave}}, \ and\
  \bibinfo {author} {\bibfnamefont {K.}~\bibnamefont {Burke}},\ }\href
  {\doibase 10.1063/1.1651060} {\bibfield  {journal} {\bibinfo  {journal} {J.
  Chem. Phys.}\ }\textbf {\bibinfo {volume} {120}},\ \bibinfo {pages} {5932 }
  (\bibinfo {year} {2004})}\BibitemShut {NoStop}%
\bibitem [{\citenamefont {Helbig}\ \emph {et~al.}(2011)\citenamefont {Helbig},
  \citenamefont {Fuks}, \citenamefont {Tokatly}, \citenamefont {Appel},
  \citenamefont {Gross},\ and\ \citenamefont {Rubio}}]{HFTAGR11}%
  \BibitemOpen
  \bibfield  {author} {\bibinfo {author} {\bibfnamefont {N.}~\bibnamefont
  {Helbig}}, \bibinfo {author} {\bibfnamefont {J.}~\bibnamefont {Fuks}},
  \bibinfo {author} {\bibfnamefont {I.}~\bibnamefont {Tokatly}}, \bibinfo
  {author} {\bibfnamefont {H.}~\bibnamefont {Appel}}, \bibinfo {author}
  {\bibfnamefont {E.}~\bibnamefont {Gross}}, \ and\ \bibinfo {author}
  {\bibfnamefont {A.}~\bibnamefont {Rubio}},\ }\href {\doibase
  http://dx.doi.org/10.1016/j.chemphys.2011.06.010} {\bibfield  {journal}
  {\bibinfo  {journal} {Chem. Phys.}\ }\textbf {\bibinfo {volume} {391}},\
  \bibinfo {pages} {1 } (\bibinfo {year} {2011})}\BibitemShut {NoStop}%
\bibitem [{\citenamefont {Elliott}\ \emph {et~al.}(2011)\citenamefont
  {Elliott}, \citenamefont {Goldson}, \citenamefont {Canahui},\ and\
  \citenamefont {Maitra}}]{EGCM11}%
  \BibitemOpen
  \bibfield  {author} {\bibinfo {author} {\bibfnamefont {P.}~\bibnamefont
  {Elliott}}, \bibinfo {author} {\bibfnamefont {S.}~\bibnamefont {Goldson}},
  \bibinfo {author} {\bibfnamefont {C.}~\bibnamefont {Canahui}}, \ and\
  \bibinfo {author} {\bibfnamefont {N.~T.}\ \bibnamefont {Maitra}},\ }\href
  {\doibase http://dx.doi.org/10.1016/j.chemphys.2011.03.020} {\bibfield
  {journal} {\bibinfo  {journal} {Chem. Phys.}\ }\textbf {\bibinfo {volume}
  {391}},\ \bibinfo {pages} {110 } (\bibinfo {year} {2011})}\BibitemShut
  {NoStop}%
\bibitem [{\citenamefont {Farzanehpour}\ and\ \citenamefont
  {Tokatly}(2012)}]{FarTok2012}%
  \BibitemOpen
  \bibfield  {author} {\bibinfo {author} {\bibfnamefont {M.}~\bibnamefont
  {Farzanehpour}}\ and\ \bibinfo {author} {\bibfnamefont {I.~V.}\ \bibnamefont
  {Tokatly}},\ }\href {\doibase 10.1103/PhysRevB.86.125130} {\bibfield
  {journal} {\bibinfo  {journal} {Phys. Rev. B}\ }\textbf {\bibinfo {volume}
  {86}},\ \bibinfo {pages} {125130} (\bibinfo {year} {2012})}\BibitemShut
  {NoStop}%
\bibitem [{\citenamefont {Elliott}\ \emph {et~al.}(2012)\citenamefont
  {Elliott}, \citenamefont {Fuks}, \citenamefont {Rubio},\ and\ \citenamefont
  {Maitra}}]{EFRM12}%
  \BibitemOpen
  \bibfield  {author} {\bibinfo {author} {\bibfnamefont {P.}~\bibnamefont
  {Elliott}}, \bibinfo {author} {\bibfnamefont {J.~I.}\ \bibnamefont {Fuks}},
  \bibinfo {author} {\bibfnamefont {A.}~\bibnamefont {Rubio}}, \ and\ \bibinfo
  {author} {\bibfnamefont {N.~T.}\ \bibnamefont {Maitra}},\ }\href {\doibase
  10.1103/PhysRevLett.109.266404} {\bibfield  {journal} {\bibinfo  {journal}
  {Phys. Rev. Lett.}\ }\textbf {\bibinfo {volume} {109}},\ \bibinfo {pages}
  {266404} (\bibinfo {year} {2012})}\BibitemShut {NoStop}%
\bibitem [{\citenamefont {Maitra}\ \emph {et~al.}(2002)\citenamefont {Maitra},
  \citenamefont {Burke},\ and\ \citenamefont {Woodward}}]{MBW02}%
  \BibitemOpen
  \bibfield  {author} {\bibinfo {author} {\bibfnamefont {N.~T.}\ \bibnamefont
  {Maitra}}, \bibinfo {author} {\bibfnamefont {K.}~\bibnamefont {Burke}}, \
  and\ \bibinfo {author} {\bibfnamefont {C.}~\bibnamefont {Woodward}},\
  }\href@noop {} {\bibfield  {journal} {\bibinfo  {journal} {Phys. Rev. Lett.}\
  }\textbf {\bibinfo {volume} {89}},\ \bibinfo {pages} {023002} (\bibinfo
  {year} {2002})}\BibitemShut {NoStop}%
\bibitem [{\citenamefont {Elliott}\ and\ \citenamefont {Maitra}(2012)}]{EM12}%
  \BibitemOpen
  \bibfield  {author} {\bibinfo {author} {\bibfnamefont {P.}~\bibnamefont
  {Elliott}}\ and\ \bibinfo {author} {\bibfnamefont {N.~T.}\ \bibnamefont
  {Maitra}},\ }\href {\doibase 10.1103/PhysRevA.85.052510} {\bibfield
  {journal} {\bibinfo  {journal} {Phys. Rev. A}\ }\textbf {\bibinfo {volume}
  {85}},\ \bibinfo {pages} {052510} (\bibinfo {year} {2012})}\BibitemShut
  {NoStop}%
\bibitem [{\citenamefont {Ruggenthaler}\ \emph {et~al.}(2013)\citenamefont
  {Ruggenthaler}, \citenamefont {Nielsen},\ and\ \citenamefont {van
  Leeuwen}}]{RNvL13}%
  \BibitemOpen
  \bibfield  {author} {\bibinfo {author} {\bibfnamefont {M.}~\bibnamefont
  {Ruggenthaler}}, \bibinfo {author} {\bibfnamefont {S.~E.~B.}\ \bibnamefont
  {Nielsen}}, \ and\ \bibinfo {author} {\bibfnamefont {R.}~\bibnamefont {van
  Leeuwen}},\ }\href {\doibase 10.1103/PhysRevA.88.022512} {\bibfield
  {journal} {\bibinfo  {journal} {Phys. Rev. A}\ }\textbf {\bibinfo {volume}
  {88}},\ \bibinfo {pages} {022512} (\bibinfo {year} {2013})}\BibitemShut
  {NoStop}%
\bibitem [{\citenamefont {Ullrich}(2012)}]{Ullrich:12}%
  \BibitemOpen
  \bibfield  {author} {\bibinfo {author} {\bibfnamefont {C.}~\bibnamefont
  {Ullrich}},\ }\href@noop {} {\emph {\bibinfo {title} {Time-Dependent
  Density-Functional Theory}}}\ (\bibinfo  {publisher} {Oxford University
  Press},\ \bibinfo {address} {Oxford},\ \bibinfo {year} {2012})\BibitemShut
  {NoStop}%
\bibitem [{\citenamefont {Marques}\ and\ \citenamefont {Rubio}(2009)}]{KK09}%
  \BibitemOpen
  \bibfield  {author} {\bibinfo {author} {\bibfnamefont {M.}~\bibnamefont
  {Marques}}\ and\ \bibinfo {author} {\bibfnamefont {A.}~\bibnamefont
  {Rubio}},\ }\href {\doibase 10.1039/B908105B} {\bibfield  {journal} {\bibinfo
   {journal} {Phys. Chem. Chem. Phys.}\ }\textbf {\bibinfo {volume} {11}},\
  \bibinfo {pages} {4436} (\bibinfo {year} {2009})}\BibitemShut {NoStop}%
\bibitem [{BKK()}]{BKK11}%
  \BibitemOpen
  \href@noop {} {\enquote {\bibinfo {title} {See the special issue of chemical
  physics, 2011, {\bf 391} (1), guest-edited by {R. B}aer, {L. K}ronik, {S.
  K}{\"u}mmel.}}\ }\BibitemShut {NoStop}%
\bibitem [{\citenamefont {Fuks}\ \emph {et~al.}(2011)\citenamefont {Fuks},
  \citenamefont {Helbig}, \citenamefont {Tokatly},\ and\ \citenamefont
  {Rubio}}]{FHTR11}%
  \BibitemOpen
  \bibfield  {author} {\bibinfo {author} {\bibfnamefont {J.~I.}\ \bibnamefont
  {Fuks}}, \bibinfo {author} {\bibfnamefont {N.}~\bibnamefont {Helbig}},
  \bibinfo {author} {\bibfnamefont {I.~V.}\ \bibnamefont {Tokatly}}, \ and\
  \bibinfo {author} {\bibfnamefont {A.}~\bibnamefont {Rubio}},\ }\href
  {\doibase 10.1103/PhysRevB.84.075107} {\bibfield  {journal} {\bibinfo
  {journal} {Phys. Rev. B}\ }\textbf {\bibinfo {volume} {84}},\ \bibinfo
  {pages} {075107} (\bibinfo {year} {2011})}\BibitemShut {NoStop}%
\bibitem [{\citenamefont {Fuks}\ \emph {et~al.}(2013)\citenamefont {Fuks},
  \citenamefont {Elliott}, \citenamefont {Rubio},\ and\ \citenamefont
  {Maitra}}]{FERM13}%
  \BibitemOpen
  \bibfield  {author} {\bibinfo {author} {\bibfnamefont {J.~I.}\ \bibnamefont
  {Fuks}}, \bibinfo {author} {\bibfnamefont {P.}~\bibnamefont {Elliott}},
  \bibinfo {author} {\bibfnamefont {A.}~\bibnamefont {Rubio}}, \ and\ \bibinfo
  {author} {\bibfnamefont {N.~T.}\ \bibnamefont {Maitra}},\ }\href {\doibase
  10.1021/jz302099f} {\bibfield  {journal} {\bibinfo  {journal} {J. Phys. Chem.
  Lett.}\ }\textbf {\bibinfo {volume} {4}},\ \bibinfo {pages} {735} (\bibinfo
  {year} {2013})}\BibitemShut {NoStop}%
\bibitem [{\citenamefont {Thiele}\ \emph {et~al.}(2008)\citenamefont {Thiele},
  \citenamefont {Gross},\ and\ \citenamefont {K\"ummel}}]{TGK08}%
  \BibitemOpen
  \bibfield  {author} {\bibinfo {author} {\bibfnamefont {M.}~\bibnamefont
  {Thiele}}, \bibinfo {author} {\bibfnamefont {E.~K.~U.}\ \bibnamefont
  {Gross}}, \ and\ \bibinfo {author} {\bibfnamefont {S.}~\bibnamefont
  {K\"ummel}},\ }\href {\doibase 10.1103/PhysRevLett.100.153004} {\bibfield
  {journal} {\bibinfo  {journal} {Phys. Rev. Lett.}\ }\textbf {\bibinfo
  {volume} {100}},\ \bibinfo {pages} {153004} (\bibinfo {year}
  {2008})}\BibitemShut {NoStop}%
\bibitem [{\citenamefont {Ramsden}\ and\ \citenamefont {Godby}(2012)}]{RG12}%
  \BibitemOpen
  \bibfield  {author} {\bibinfo {author} {\bibfnamefont {J.~D.}\ \bibnamefont
  {Ramsden}}\ and\ \bibinfo {author} {\bibfnamefont {R.~W.}\ \bibnamefont
  {Godby}},\ }\href {\doibase 10.1103/PhysRevLett.109.036402} {\bibfield
  {journal} {\bibinfo  {journal} {Phys. Rev. Lett.}\ }\textbf {\bibinfo
  {volume} {109}},\ \bibinfo {pages} {036402} (\bibinfo {year}
  {2012})}\BibitemShut {NoStop}%
\bibitem [{\citenamefont {Carrascal}\ and\ \citenamefont
  {Ferrer}(2012)}]{CF12}%
  \BibitemOpen
  \bibfield  {author} {\bibinfo {author} {\bibfnamefont {D.~J.}\ \bibnamefont
  {Carrascal}}\ and\ \bibinfo {author} {\bibfnamefont {J.}~\bibnamefont
  {Ferrer}},\ }\href {\doibase 10.1103/PhysRevB.85.045110} {\bibfield
  {journal} {\bibinfo  {journal} {Phys. Rev. B}\ }\textbf {\bibinfo {volume}
  {85}},\ \bibinfo {pages} {045110} (\bibinfo {year} {2012})}\BibitemShut
  {NoStop}%
\bibitem [{\citenamefont {Ullrich}(2006)}]{U06}%
  \BibitemOpen
  \bibfield  {author} {\bibinfo {author} {\bibfnamefont {C.~A.}\ \bibnamefont
  {Ullrich}},\ }\href@noop {} {\bibfield  {journal} {\bibinfo  {journal} {J.
  Chem. Phys.}\ }\textbf {\bibinfo {volume} {125}},\ \bibinfo {pages} {234108}
  (\bibinfo {year} {2006})}\BibitemShut {NoStop}%
\bibitem [{\citenamefont {Verdozzi}(2008)}]{V08}%
  \BibitemOpen
  \bibfield  {author} {\bibinfo {author} {\bibfnamefont {C.}~\bibnamefont
  {Verdozzi}},\ }\href {\doibase 10.1103/PhysRevLett.101.166401} {\bibfield
  {journal} {\bibinfo  {journal} {Phys. Rev. Lett.}\ }\textbf {\bibinfo
  {volume} {101}},\ \bibinfo {pages} {166401} (\bibinfo {year}
  {2008})}\BibitemShut {NoStop}%
\bibitem [{\citenamefont {Levy}(1979)}]{Levy:79}%
  \BibitemOpen
  \bibfield  {author} {\bibinfo {author} {\bibfnamefont {M.}~\bibnamefont
  {Levy}},\ }\href@noop {} {\bibfield  {journal} {\bibinfo  {journal}
  {Proc.~Natl.~Acad.~Sci.~USA}\ }\textbf {\bibinfo {volume} {76}},\ \bibinfo
  {pages} {6062} (\bibinfo {year} {1979})}\BibitemShut {NoStop}%
\bibitem [{\citenamefont {Levy}(1982)}]{Levy82}%
  \BibitemOpen
  \bibfield  {author} {\bibinfo {author} {\bibfnamefont {M.}~\bibnamefont
  {Levy}},\ }\href {\doibase 10.1103/PhysRevA.26.1200} {\bibfield  {journal}
  {\bibinfo  {journal} {Phys. Rev. A}\ }\textbf {\bibinfo {volume} {26}},\
  \bibinfo {pages} {1200} (\bibinfo {year} {1982})}\BibitemShut {NoStop}%
\bibitem [{\citenamefont {Lieb}(1983)}]{Lieb:83}%
  \BibitemOpen
  \bibfield  {author} {\bibinfo {author} {\bibfnamefont {E.}~\bibnamefont
  {Lieb}},\ }\href@noop {} {\bibfield  {journal} {\bibinfo  {journal}
  {Int.~J.~Quantum Chem.}\ }\textbf {\bibinfo {volume} {24}},\ \bibinfo {pages}
  {243} (\bibinfo {year} {1983})}\BibitemShut {NoStop}%
\bibitem [{\citenamefont {Li}\ and\ \citenamefont {Ullrich}(2008)}]{LiUll2008}%
  \BibitemOpen
  \bibfield  {author} {\bibinfo {author} {\bibfnamefont {Y.}~\bibnamefont
  {Li}}\ and\ \bibinfo {author} {\bibfnamefont {C.~A.}\ \bibnamefont
  {Ullrich}},\ }\href {\doibase 10.1063/1.2955733} {\bibfield  {journal}
  {\bibinfo  {journal} {J. Chem. Phys.}\ }\textbf {\bibinfo {volume} {129}},\
  \bibinfo {eid} {044105} (\bibinfo {year} {2008})}\BibitemShut {NoStop}%
\bibitem [{\citenamefont {Ruggenthaler}\ and\ \citenamefont
  {Bauer}(2009)}]{RB09}%
  \BibitemOpen
  \bibfield  {author} {\bibinfo {author} {\bibfnamefont {M.}~\bibnamefont
  {Ruggenthaler}}\ and\ \bibinfo {author} {\bibfnamefont {D.}~\bibnamefont
  {Bauer}},\ }\href@noop {} {\bibfield  {journal} {\bibinfo  {journal} {Phys.
  Rev. Lett.}\ }\textbf {\bibinfo {volume} {102}},\ \bibinfo {pages} {233001}
  (\bibinfo {year} {2009})}\BibitemShut {NoStop}%
\bibitem [{Note1()}]{Note1}%
  \BibitemOpen
  \bibinfo {note} {Since the whole right hand side in this equation is
  proportional to the time derivative of the denominator, turning the square
  root into zero does not lead to a divergence because the numerator vanishes
  at the same time. Also, for any physical evolution the expression under the
  square root always stays nonnegative as $|\protect \mathaccentV {dot}05Fd|$
  for a lattice is properly bounded from above ~\protect \cite
  {V08,TokatlyL2011,FarTok2012}.}\BibitemShut {Stop}%
\bibitem [{\citenamefont {Maitra}(2011)}]{MaitraTDDFT2011}%
  \BibitemOpen
  \bibfield  {author} {\bibinfo {author} {\bibfnamefont {N.~T.}\ \bibnamefont
  {Maitra}},\ }in\ \href@noop {} {\emph {\bibinfo {booktitle} {Time-Dependent
  Density Functional Theory}}},\ \bibinfo {series} {Lecture Notes in physics},
  Vol.\ \bibinfo {volume} {706}\ (\bibinfo  {publisher} {Springer},\ \bibinfo
  {address} {Heidelberg},\ \bibinfo {year} {2011})\BibitemShut {NoStop}%
\bibitem [{\citenamefont {Giuliani}\ and\ \citenamefont
  {Vignale}(2005)}]{VignaleBook}%
  \BibitemOpen
  \bibfield  {author} {\bibinfo {author} {\bibfnamefont {G.~F.}\ \bibnamefont
  {Giuliani}}\ and\ \bibinfo {author} {\bibfnamefont {G.}~\bibnamefont
  {Vignale}},\ }\href@noop {} {\emph {\bibinfo {title} {Quantum theory of the
  electron liquid}}}\ (\bibinfo  {publisher} {Cambridge University Press},\
  \bibinfo {address} {Cambridge},\ \bibinfo {year} {2005})\BibitemShut
  {NoStop}%
\bibitem [{\citenamefont {Tokatly}(2011)}]{TokatlyL2011}%
  \BibitemOpen
  \bibfield  {author} {\bibinfo {author} {\bibfnamefont {I.~V.}\ \bibnamefont
  {Tokatly}},\ }\href {\doibase 10.1103/PhysRevB.83.035127} {\bibfield
  {journal} {\bibinfo  {journal} {Phys. Rev. B}\ }\textbf {\bibinfo {volume}
  {83}},\ \bibinfo {pages} {035127} (\bibinfo {year} {2011})}\BibitemShut
  {NoStop}%
\end{thebibliography}%

\end{document}